\documentclass{esub2acm}

\usepackage{graphicx}
\usepackage{amssymb,amsmath}
\usepackage{times}
\usepackage{mathptm}

\newtheorem{theorem}{Theorem}
\newtheorem{proposition}{Proposition}
\newtheorem{remark}{Remark}
\newtheorem{fact}{Fact}

\begin{document}

\title{Routing Permutations\\ in Partitioned Optical Passive Stars Networks}

\author{Alessandro Mei\\
Department of Computer Science, University of Rome
``La Sapienza'', Italy, \and
Romeo Rizzi\\
Department of Mathematics, University of Trento, Italy.}

\begin{abstract}
It is shown that a Partitioned Optical Passive Stars (POPS) network
with $g$ groups and $d$ processors per group can route
any permutation among the $n=dg$ processors in one slot when $d=1$
and $2\lceil d/g\rceil$ slots when $d>1$. The number of slots used is
optimal in the worst case, and is at most the double of the optimum
for all permutations $\pi$ such that $\pi(i)\neq i$, for all $i$. 
\end{abstract}


\markboth{A. Mei and R. Rizzi}
     {Routing Permutations in POPS Networks}

\maketitle

\section{Introduction}

\begin{figure}
\begin{center}
\includegraphics{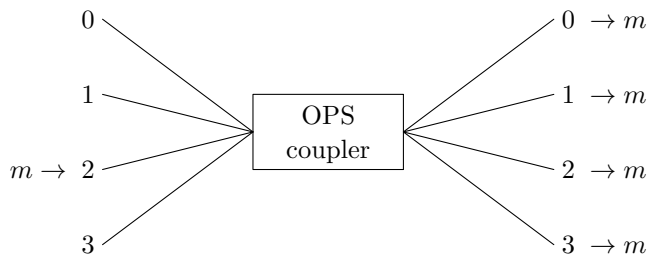}
\end{center}
\caption{A $4\times 4$ Optical Passive Star (OPS) coupler.}
\label{fig:opscoupler}
\end{figure}

The Partitioned Optical Passive Star (POPS) network
\cite{clmtg94,gmclt95,gm98,mgcl98} is a SIMD
interconnection network that uses multiple optical passive star
(OPS) couplers. 
A $d\times d$ OPS coupler (see Figure \ref{fig:opscoupler}) is
an all-optical passive device which is capable of
receiving an optical signal from one of its $d$ sources and broadcast it to
all of its $d$ destinations. Being a passive all-optical technology
it benefits from a number of characteristics such as no opto-electronic
conversion, high noise immunity, and low latency.

The number of processors of the network is denoted by $n$, and each processor
has a distinct index in $\{0,\dotsc, n-1\}$.
The $n$ processors are partitioned into $g=n/d$
groups in such a way that processor $i$ belongs to group
$\mathrm{group}(i):=\lfloor i/d\rfloor$. It is assumed that $d$ divides
$n$, consequently, each group consists of $d$ processors.
For each pair of groups $a,b\in\{0,\dotsc,g-1\}$, a coupler $c(b,a)$ is
introduced which has all the processors of group $a$ as sources and all
the processors of group $b$ as destinations. The number of couplers
used is $g^2$. 
Such an architecture will be denoted by $\mathrm{POPS}(d,g)$ (see
Figure \ref{fig:pops}).
\begin{figure}
\begin{center}
\includegraphics{pops2.mps}
\end{center}
\caption{A $\mathrm{POPS}(3,2)$.}
\label{fig:pops}
\end{figure}

For all $i\in\{0,\dotsc,n-1\}$, processor $i$ has $g$ transmitters which are
connected to couplers $c(a,\mathrm{group}(i))$, $a=0,\dotsc,g-1$.
Similarly, processor $i$ has $g$ receivers connected to couplers
$c(\mathrm{group}(i),b)$, $b=0,\dotsc,g-1$. During a step of
computation, each processor in parallel:
\begin{itemize}
\item
Performs some local computations;
\item
sends a packet to a subset of its transmitters;
\item
receives a packet from one of its receivers.
\end{itemize}
In order to avoid conflicts, there shouldn't be any pair of processors
sending a packet to the same coupler. The time needed to perform such a step
is referred to as a \emph{slot}.

One of the advantages of a $\mathrm{POPS}(d,g)$ network is
that its diameter is 1. A packet can
be sent from processor $i$ to processor $j$, $i\neq j$, in one slot
by using coupler $c(\mathrm{group}(j),\mathrm{group}(i))$. However,
its bandwidth varies according to $g$. In a $\mathrm{POPS}(n,1)$ network,
only one packet can be sent through the single coupler per slot.
On the other extreme, a $\mathrm{POPS}(1,n)$ network is a highly expensive,
fully interconnected optical network using $n^2$ OPS couplers.

A one-to-all communication pattern can also be performed in only one slot in
the following way: Processor $i$ (the speaker) sends the packet to
all the couplers $c(a,\mathrm{group}(i))$, $a\in\{0,\dotsc,g-1\}$,
during the same slot all the processors $j$, $j\in\{0,\dotsc,n-1\}$,
can receive the packet trough coupler
$c(\mathrm{group}(j),\mathrm{group}(i))$.

The POPS network model has been used to develop a number of non trivial
algorithms. Several common communication patterns are realized in
\cite{gm98}. Simulation algorithms for the mesh and hypercube interconnection
networks can be found in \cite{s00a}. Algorithms for data sum, prefix
sum, consecutive sum, adjacent sum, and several data movement operations
are also described in \cite{s00a}. An algorithm for matrix multiplication 
is provided in \cite{s00b}.
These algorithms are based on sophisticated communication patterns,
which have been investigated one by one, and shown to be routable on a
$\mathrm{POPS}(d,g)$ network. However, most of these patterns belong to a
more general class of permutation routing problems whose routability on the
POPS network was not known in general. In this paper,
we show that a $\mathrm{POPS}(d,g)$ network can efficiently route
$n=dg$ packets arranged in the $n$ processors according to any permutation,
generalizing and unifying several known results appeared in the recent
literature.

\section{Definition of the Problem and Related Work}

Let $\mathbb{N}_n := \{0,1, \ldots, n-1\}$
denote the set of the first $n$ natural numbers, and let $\pi$ be
a permutation of the set $\mathbb{N}_n$.
A \emph{permutation routing problem} consists of a set of $n$ packets
$p_0,\ldots, p_{n-1}$. Packet $p_i$ is stored in the local
memory of processor~$i$, for all $i\in\mathbb{N}_n$, and has a
desired destination $\pi(i)$. The problem is to route the packets
to their destinations in as few slots as possible.

No general solution has been given for this problem on the POPS network.
Efficient routings are known for a few particular permutations, which
have been independently attacked, and most of
them require one slot when $d=1$ and $2\lceil d/g\rceil$ slots when $d>1$.
Here follow a few examples.

In \cite{gm98}, a characterization is given of the permutation
routing problems that can be routed in a single slot. However, only a
very restricted number of permutations fall in this class. Indeed, if
two packets originating at the same group are to be routed
to the same destination group, then one slot is obviously not enough to route
all the packets.

In \cite{s00a}, several permutation routing problems are considered in the
context of the simulation of hypercube and mesh-connected computers on
the POPS network. Assume that processor~$i$ of an $n=2^D$ processor SIMD
hypercube is mapped onto processor~$i$ of a $\mathrm{POPS}(d,g)$ network,
$dg=n$. For every fixed $b$, $0\le b<D$, a primitive communication pattern
is defined such that processor~$i$ sends a packet to processor~$i^{(b)}$,
where $i^{(b)}$ is the number whose binary representation differs from that
of $i$ only in bit $b$. Each of the $D$ communication patterns defined is
a permutation routing problem. Theorem~1 of \cite{s00a} shows that all
of them can be routed in one slot when $d=1$
and $2\lceil d/g\rceil$ slots when $d>1$.
 
The same result has been obtained when considering the problem of simulating
an $N\times N$ SIMD mesh with wraparound, where data can be moved one processor
up/down along the columns of the mesh, or right/left along the rows
of the mesh. Again, assuming that processor~$(i,j)$ of the mesh is mapped
onto processor~$i+jN$ of a $\mathrm{POPS}(d,g)$ network ($dg=N^2$ and
either $d$ or $g$
divides $N$), Theorem~2 of \cite{s00a} shows that one slot when $d=1$
and $2\lceil d/g\rceil$ slots when $d>1$ are enough to route each of the
four permutation routing problems.
 
The routability of other specific permutation routing problems is investigated
in \cite{s00b}. For example, a vector reversal (a permutation routing
problem, where $\pi(i)=n-1-i$, $0\le i<n$) is shown
to be routable in one slot when $d=1$
and $2\lceil d/g\rceil$ slots when $d>1$ on a $\mathrm{POPS}(d,g)$ network,
$dg=n$, which is optimal when $g$ is even.
To route a matrix transpose, conversely, $\lceil d/g\rceil$ is
the optimal number of slots required.

Moreover, \cite{s00b} considers
BPC permutations. A BPC permutation is a rearrangement of the bits of the
source processor index, while some or all of the bits can be complemented.
Formally, assume that $n$ is a power of 2, $n=2^k$, and that
the binary representation of $i$ is $\left[i_{k-1}i_{k-2}\cdots i_0\right]_2$,
the set of BPC permutations is the smallest set $\mathrm{BPC}$ closed
under composition such that:
\begin{enumerate}
\item
$\pi(i)=\left[i_{\sigma(k-1)}i_{\sigma(k-2)}\cdots i_{\sigma(0)}\right]_2\in
\mathrm{BPC}$, for all $\sigma$ permutation of $\mathbb{N}_k$;
\item
$\pi(i)=\left[i_{k-1}\cdots \overline{i_{j}}\cdots i_0\right]_2\in\mathrm{BPC}$,
for all $j$.
\end{enumerate}
Again, \cite{s00b} describes how BPC permutations can be routed
in one slot when $d=1$ and $2\lceil d/g\rceil$ slots when $d>1$ on a
$\mathrm{POPS}(d,g)$ network, $dg=n$.

In this paper we unify, generalize, and simplify the previously known results,
by showing that a $\mathrm{POPS}(d,g)$ network, $dg=n$, can route
\emph{any} permutation in one slot when $d=1$ and $2\lceil d/g\rceil$ slots
when $d>1$. This gives evidence of the
versatility of the network. For example, a consequence of our
Theorem~\ref{thm:main} is that the simulation results
for hypercube and mesh-connected computers shown in \cite{s00a} do not depend
on how the processors of the simulated architecture are mapped
onto the processors of the POPS network, provided that it is a one-to-one
mapping, which is somewhat surprising.

\section{Routing Permutations in the POPS network}

Assume the permutation routing problem defined by $\pi$ on a
$\mathrm{POPS}(d,g)$ network, $dg=n$, where $\pi$ is a permutation 
of $N_n$. Our goal is to prove that $\pi$ can be routed in one slot when
$d=1$ and $2\lceil d/g\rceil$ slots when $d>1$.

\begin{figure}
\begin{center}
\includegraphics{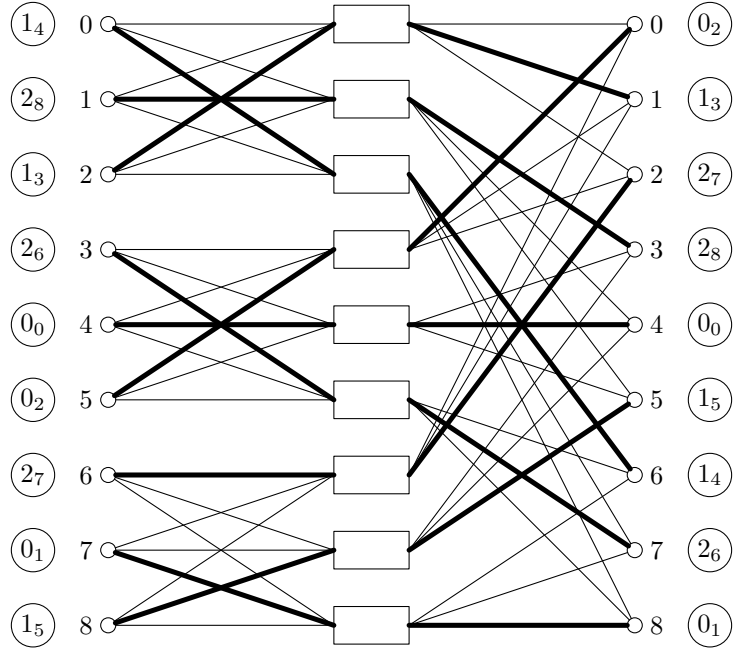}
\end{center}
\caption{Getting to a fair distribution on a $\mathrm{POPS}(3,3)$.
Packets are drawn as circles
next to their sources on the left. Inside each packet its destination $x_y$
can be found, where $y$ is the index of the destination processor, and $x$
is its group. On the right, the intermediate destination of the packet as
described by Section \ref{sec:slotone}.}
\label{fig:slotone}
\end{figure}

We start, for the ease of explanation, from the case $d=g=\sqrt{n}$.
In this case, for most permutations one slot is not enough to route 
all the packets to destination. Take, as an example, the permutation
shown in Figure~\ref{fig:slotone}. Packets starting from processor~4
and processor~5, both belonging to group 1, have the same group 0
as desired destination. If only one slot is allowed,
there is an unavoidable conflict on coupler $\mathrm{c}(0,1)$.
Hence, two slots are necessary to route $\pi$.

It is not hard to find a sufficient condition for a set of packets
to be routable in one slot.
We will say that $m$ packets, each with a different destination, are
arranged according to a
\emph{fair distribution} in a $\mathrm{POPS}(d,g)$ network if no two packets
are stored in the same processor, and no two packets with the same destination
group are stored in the same group. In this case, we will also say that
the packets are \emph{fairly distributed}.

It is straightforward to see that a fairly
distributed set of packets is routable in one slot. Indeed, no conflict
occurs on any coupler.
\begin{fact}
\label{fact:phasetwo}
In a $\mathrm{POPS}(d,g)$ network, a fairly distributed set of $m$ packets
can be routed to destination in one slot.
\end{fact}
When $d=g=\sqrt{n}$, only a very small number of permutations can be routed in
one slot. However, we will show that all of them can be routed in two slots.
The idea is that one slot is always enough to move a set of $n$ packets
arranged according to $\pi$ in such a way to become fairly distributed. Then,
a second one routes all the packets to destination by Fact~\ref{fact:phasetwo}.

Next, in Subsection~\ref{sec:slotone}, we formalize the above intuition, and
demonstrate our claim, properly generalized in order to deal with any
value of $d$ and $g$. Note that, for a set of packets to be fairly
distributed, we don't really need to care about their processor destination.
What we need is just to know what \emph{group} destination
each packet has. Thus, in Subsection~\ref{sec:slotone} we can reduce our
discussion to source groups and destination groups. $d$ packets originate at
each source group, and $d$ packets have a specific destination group.
 
\subsection{Permutation Routing: Getting to a Fair Distribution}
\label{sec:slotone}

A {\em list system} is a triple $(S,T,\mathcal{L})$,
where $S$ is a set of $n_1 := |S|$ {\em source nodes},
$T$ is a set of $n_2 := |T|$ {\em target nodes},
and $\mathcal{L}:S\times \mathbb{N}_{\Delta_1} \mapsto S$
assigns a list $L_s$ of $\Delta_1\le n_2$ not necessarily distinct
elements from $S$ to every source node $s\in S$.
We also let $l(s,s')$ specify how many times
the element $s'\in S$ appears into list $L_s$.
A list system is called {\em proper} when $n_2$ divides 
$n_1\Delta_1$, and $\sum_{s\in S} l(s,s') = \Delta_1$ for every $s'\in S$.

Let $\Delta_2 := \frac{n_1 \Delta_1}{n_2}$. A {\em fair distribution} is an
assignment $f: S\times \mathbb{N}_{\Delta_1} \mapsto T$
such that
\begin{align}
\label{eqn:prop1}
   & |\{f(s,i) \; | \; i\in \mathbb{N}_{\Delta_1}\}| = \Delta_1
         \;\textrm{for every $s\in S$;}\\
\label{eqn:prop2}
   & |\{(s,i)\in S \times \mathbb{N}_{\Delta_1} \; | \; f(s,i)=t\}| = \Delta_2
         \;\textrm{for every $t\in T$;}\\
\label{eqn:prop3}
\begin{split}
& \textrm{if}\; (s_1,i_1)\neq (s_2,i_2)\; \textrm{and} \;
\mathcal{L}(s_1,i_1)= \mathcal{L}(s_2,i_2),\;\textrm{then}\;
f(s_1,i_1)\neq f(s_2,i_2),\\
& \indent\textrm{for every $s_1, s_2\in S$ and every $i_1, i_2\in \Delta_1$.}
\end{split}
\end{align}

\begin{theorem}
\label{thm:slotone}
   Every proper list system admits a fair distribution.
\end{theorem}
\begin{proof}
   Let $S':= \{s' \, | \, s\in S\}$.
   Consider the bipartite multigraph $G=(S,S';E)$,
   on node classes $S$ and $S'$,
   and having precisely $l(s,s')$ edges with one endnode in $s$ and the
   other in $s'$.
   Clearly, for every $s\in S$,
   $E$ contains precisely $\Delta_1$ edges
   incident with $s$, namely the edges
   $\{s,\mathcal{L}(s,i)\}$ for $i\in \mathbb{N}_{\Delta_1}$.
   Moreover, for every $s'\in S$,
   $E$ contains precisely $\Delta_1$ edges
   incident with $s'$, since the list system is proper (and by (\ref{eqn:f1})).
   Our problem is to find an edge-coloring of $G$
   with $n_2$ ($\geq \Delta_1$ and such that $n_2$ divides $n_1 \Delta_1$)
   colors and such that each color class
   has size precisely $\Delta_2 := \frac{n_1 \Delta_1}{n_2}$.\\

Let $V$ be a set of $n_1-\Delta_2$ new nodes
and $V':= \{v' \, | \, v\in V\}$.
Let $H_1 = (V,S';F_1)$ be any bipartite
$(n_2,n_2-\Delta_1)$-regular bipartite graph
on node classes $V$ and $S'$.
Let $H_2 = (V',S;F_2)$ be any bipartite
$(n_2,n_2-\Delta_1)$-regular bipartite graph
on node classes $V'$ and $S$.
Consider the bipartite
$n_2$-regular multigraph
$\overline{G}=(S\cup V,S'\cup V'; E \cup F_1 \cup F_2)$.
   By K\"onig's theorem~\cite{konig1,konig2},
   we can edge-color $\overline{G}$ with $n_2$ colors,
   that is,
   we can decompose $E \cup F_1 \cup F_2$
   into $n_2$ \emph{perfect} matchings
   $M_1, \ldots, M_{n_2}$ of $\overline{G}$.
   We propose
   $M_1 \setminus F_1 \setminus F_2,
      \ldots, M_{n_2} \setminus F_1 \setminus F_2$
   as the required edge-coloring of $G$.
   Indeed, $M_1 \setminus F_1 \setminus F_2,
            \ldots, M_{n_2} \setminus F_1 \setminus F_2$
   is a decomposition of $E$ into $n_2$ matchings of $G$
   and $|M_i \setminus F_1 \setminus F_2| =
    |M_i|-(|V|+|V'|) =
    (n_1 + |V|) -2|V| =
    n_1 - |V| = n_1 - n_1 + \Delta_2 = \Delta_2$,
   for every $i=1, \ldots, \Delta$.
\end{proof}

\begin{remark}
\label{rem:remark}
   The above proof is algorithmic.
   The computational bottleneck is in computing a $1$-factorization
   of a bipartite $n_2$-regular multigraph
   on $n := 4n_1-2\Delta_2$ nodes and with
   $m := nn_2$ edges.
   This can be done in
   $O(n_2 m)$
   as in~\cite{Sch}
   or in
   $O(m\log n_2 + \frac {m}{ n_2}
    \log \frac {m}{n_2} \log n_2)$
   as in~\cite{Rajai}
   and in virtue of the algorithm described in~\cite{Rfactor}.
\end{remark}

\subsection{Permutation Routing: the Main Theorem}

The following theorem describes our main result. Note that the
routing found by Theorem \ref{thm:main} has the property that at each step
of computation each processor stores exactly one packet.

\begin{theorem}
\label{thm:main}
A $\mathrm{POPS}(d,g)$ network can route any permutation $\pi$ among the $n=dg$
processors using one slot when $d=1$ and $2\lceil d/g\rceil$ slots when $d>1$.
\end{theorem}

\begin{proof}
When $d=1$, a $\mathrm{POPS}(1,n)$ network is equivalent to an $n$ processor
clique, the network is fully interconnected, and the claim of the theorem is
thus trivial.

Now, consider the case when $1<d\le g$. We will show that $\pi$ can be
routed in $2\lceil d/g\rceil=2$ slots. Take the list system
$(\mathbb{N}_g,\mathbb{N}_g,\mathcal{L})$,
where $\mathcal{L}:\mathbb{N}_g\times \mathbb{N}_d\mapsto \mathbb{N}_g$
is such that $\mathcal{L}(h,i)=\mathrm{group}(\pi(i+hd))$,
$h\in\mathbb{N}_g, i\in\mathbb{N}_d$. The list system is proper, since
$\pi$ is a permutation, and $g$ clearly divides $gd$.
By Theorem~\ref{thm:slotone},
$(\mathbb{N}_g,\mathbb{N}_g,\mathcal{L})$ admits a fair distribution
$f:\mathbb{N}_g\times \mathbb{N}_d\mapsto \mathbb{N}_g$. Consequently, $f$
maps every pair $(h,i)$ to an integer from  $\mathbb{N}_g$ in such
a way that:
\begin{align}
\label{eqn:f1}
   & |\{f(h,i) \; | \; i\in \mathbb{N}_d\}| = d
         \;\textrm{for every $h\in \mathbb{N}_g$;}\\
\label{eqn:f2}
   & |\{(h,i)\in \mathbb{N}_g \times \mathbb{N}_d \; | \; f(h,i)=j\}| = d
         \;\textrm{for every $j\in \mathbb{N}_g$;}\\
\label{eqn:f3}
\begin{split}
& \textrm{if}\; (h_1,i_1)\neq (h_2,i_2)\; \textrm{and} \;
\mathcal{L}(h_1,i_1)= \mathcal{L}(h_2,i_2),\;\textrm{then}\;
f(h_1,i_1)\neq f(h_2,i_2),\\
& \indent\textrm{for every $h_1, h_2\in \mathbb{N}_g$ and every
$i_1, i_2\in \mathbb{N}_d$.}
\end{split}
\end{align}
Permutation $\pi$ is routed in two slots. During the first slot,
$n$ packets are routed through $n$ of the $g^2$ couplers of the POPS network,
and, precisely, the packet originating at processor $i+hd$ is sent
through coupler $c(f(h,i),h)$, $h\in\mathbb{N}_g, i\in\mathbb{N}_d$.
No conflict can occur on any coupler by equation~(\ref{eqn:f1}).
Moreover, exactly $d$ packets arrive at group $h$ by equation~(\ref{eqn:f2}),
hence, it is easy to assign a distinct processor to read each of the incoming
packets. After the first slot, the $n$ packets are fairly distributed
by equation~(\ref{eqn:f3}). Consequently, a second slot is enough to
route all of them to destination by Fact~\ref{fact:phasetwo}.

Finally, consider the case when $d>g$.
Take the list system
$(\mathbb{N}_g,\mathbb{N}_d,\mathcal{L})$,
where $\mathcal{L}:\mathbb{N}_g\times \mathbb{N}_d\mapsto \mathbb{N}_g$
is such that $\mathcal{L}(h,i)=\mathrm{group}(\pi(i+hd))$,
$h\in\mathbb{N}_g, i\in\mathbb{N}_d$. The list system is proper, since
$\pi$ is a permutation, and $d$ clearly divides $gd$.
By Theorem~\ref{thm:slotone},
$(\mathbb{N}_g,\mathbb{N}_d,\mathcal{L})$ admits a fair distribution
$f:\mathbb{N}_g\times \mathbb{N}_d\mapsto \mathbb{N}_d$. Consequently, $f$
maps every pair $(h,i)$ to an integer from  $\mathbb{N}_g$ in such
a way that equation~(\ref{eqn:f1}), equation~(\ref{eqn:f3}), and
the following equation~(\ref{eqn:g2}) hold.
\begin{equation}
\label{eqn:g2}
 |\{(h,i)\in \mathbb{N}_g \times \mathbb{N}_d \; | \; f(h,i)=j\}| = d
         \;\textrm{for every $j\in \mathbb{N}_d$.}
\end{equation}
Permutation $\pi$ is routed in
$\lceil d/g\rceil$ rounds. Each round $k$, $k=0,\ldots,\lceil d/g\rceil-1$,
consists of two slots.
During the first slot of all rounds but the last one, $g^2$ packets are routed
through the $g^2$ couplers of the POPS network, and, precisely,
the packet originating at processor $i+kg+hd$ is sent
through coupler $c(f(h,i+kg),h)$, $h\in\mathbb{N}_g, i\in\mathbb{N}_g$.
No conflict can occur on any coupler by equation~(\ref{eqn:f1}).
Moreover, exactly $g$ packets arrive at group $h$ by equation~(\ref{eqn:g2}),
hence, it is easy to assign a distinct processor (among the $g$ which just
sent a packet) to read each of the incoming packets. After the first slot,
the $g^2$ packets which moved are fairly distributed by equation~(\ref{eqn:f3}).
Consequently, a second slot is enough to route all of them to destination by
Fact~\ref{fact:phasetwo}. The last round is exactly identical to the previous
ones when $g$ divides $d$. Otherwise, only $g(d \mod g)$
packets are routed in a similar way. After $\lceil d/g\rceil$ rounds all
packets are correctly routed to destination.

The routing is completed after $\lceil d/g\rceil$ rounds, and each round
consists of two slots. Consequently, $\pi$ is routed using one slot when $d=1$
and $2\lceil d/g\rceil$ slots when $d>1$, as claimed.
\end{proof}

The routing described by the previous theorem can be computed efficiently.
The bottleneck consists in finding a fair distribution for the list system
described by $\pi$,
as in Theorem~\ref{thm:slotone} and Remark~\ref{rem:remark}.
It is easy to see that this can be done in
$O(g^3)$ or $O(g^2\log g)$, when $1<d\le g$, and in $O(dn)$ or
$O(n\log d)$ time, when $d>g$,
by using the algorithms in~\cite{Sch} and ~\cite{Rajai,Rfactor},
respectively.

\subsection{Optimality}

Theorem~\ref{thm:main} is not far from optimality for almost all
permutations. Indeed, if $\pi$ is such that $\pi(i)\neq i$ for all $i$,
then the routing found by Theorem~\ref{thm:main} uses at most the double of
the optimal number of slots.
\begin{proposition}
If $\pi$ is such that $\pi(i)\neq i$ for all $i$, then a $\mathrm{POPS}(d,g)$
network must use at least $\lceil d/g\rceil$ slots to route $\pi$.
\end{proposition}

\begin{proof}
Under the above assumptions, all packet destinations are different from
the source. Hence, at least one slot is needed by each packet to reach
the desired destination. Since a $\mathrm{POPS}(d,g)$ network can move
at most $g^2$ packets per slot, $\lceil n/g^2\rceil=\lceil d/g\rceil$ slots
must be used to route all the packets.
\end{proof}

Moreover, there exist permutations for which Theorem~\ref{thm:main} is optimal.
One example is vector reversal (when $g$ is even), the proof can be found
in \cite{s00b}. 
A straightforward generalization of the proof in \cite{s00b} shows
that many other permutations have the same property.
\begin{proposition}
If $\pi$ is such that $\mathrm{group}(i)\neq \mathrm{group}(\pi(i))$ and
\[
\mathrm{group}(i)=\mathrm{group}(j)\Rightarrow
\mathrm{group}(\pi(i))=\mathrm{group}(\pi(j))
\]
for all $i$ and $j$, then a $\mathrm{POPS}(d,g)$ network, $dg=n$, must use
at least $2\lceil d/g\rceil$ slots to route $\pi$.
\end{proposition}

Finally, also when the assumption that
$\mathrm{group}(i)\neq \mathrm{group}(\pi(i))$ is removed our algorithm gets
very close to an optimal number of slots.
\begin{proposition}
If $\pi$ is such that $\pi(i)\neq i$ for all $i$ and
\[
\mathrm{group}(i)=\mathrm{group}(j)\Rightarrow
\mathrm{group}(\pi(i))=\mathrm{group}(\pi(j))
\]
for all $i$ and $j$, then a $\mathrm{POPS}(d,g)$ network, $dg=n$, must use
at least $2\lceil d/(1+g)\rceil$ slots to route $\pi$.
\end{proposition}
\begin{proof}
Suppose that a $\mathrm{POPS}(d,g)$ network can route $\pi$ in $t$ slots.
If $t>d$, then it is easy to see that $t\ge 2\lceil d/(1+g)\rceil$.
Hence, we can assume without loss of generality that $t\le d$.

Since $\mathrm{group}(i)=\mathrm{group}(j)\Rightarrow
\mathrm{group}(\pi(i))=\mathrm{group}(\pi(j))$, at most $t$ packets per group
can be routed to destination in one slot only. All the other packets,
at least $d-t$ per group, have to perform at least 2 hops to get to
destination. Taking into account that a $\mathrm{POPS}(d,g)$ network can
move at most $g^2$ packets per slot, then $tg^2\ge gt+2g(d-t)$, which
implies that $t\ge 2\lceil d/(1+g)\rceil$.
\end{proof}

\section{Conclusion}

A few papers appeared in the recent literature describing how data can be moved
efficiently in a $\mathrm{POPS}(d,g)$ network. In particular, several
permutation routing problems have been independently attacked in order to show
they are routable in one slot when $d=1$ and $2\lceil d/g\rceil$ slots
when $d>1$. With Theorem~\ref{thm:main}, we demonstrate that exactly the
same result holds for any permutation $\pi$, and that the routing for $\pi$
can be efficiently computed. Moreover, the number of slots used is
optimal for
a class of permutations, and at most twice of the number of slots required
by any permutation $\pi$ such that $\pi(i)\neq i$ for all $i$.

\bibliographystyle{esub2acm}
\bibliography{p}

\end{document}